\documentclass[12pt,showpacs,prd]{revtex4}
\usepackage{epsfig,amssymb,amsfonts}

\makeatletter
\renewcommand{\@makefntext}[1]{\parindent=1em\noindent\hbox
to
1.8em{\hss$^{\@thefnmark}$}#1}
\renewcommand{\@footnotemark}{\hbox{\mathsurround=0pt$^{
\@thefnmark}$}}
\newcommand{\ftnote}[2]{\footnotemark[#1]\footnotetext[#1]{#2}}
\makeatother
\DeclareMathSymbol{\varGamma}{\mathord}{letters}{"00}
\begin{document}
\hfill{\small FZJ--IKP(TH)--2006--22}

\title{Comment on ``Once more about the $\kk$ molecule approach to the light scalars''}

\author{Yu. S. Kalashnikova}
\affiliation{Institute of Theoretical and Experimental Physics, 117218, B.Cheremushkinskaya 25, Moscow, Russia}

\author{A. E. Kudryavtsev}
\affiliation{Institute of Theoretical and Experimental Physics, 117218, B.Cheremushkinskaya 25, Moscow, Russia}

\author{A. V. Nefediev}
\affiliation{Institute of Theoretical and Experimental Physics, 117218, B.Cheremushkinskaya 25, Moscow, Russia}

\author{J. Haidenbauer}
\affiliation{Institut f\"{u}r Kernphysik, Forschungszentrum J\"{u}lich GmbH, D--52425 J\"{u}lich, Germany}

\author{C. Hanhart}
\affiliation{Institut f\"{u}r Kernphysik, Forschungszentrum J\"{u}lich GmbH, D--52425 J\"{u}lich, Germany}

\newcommand{\be}{\begin{equation}}
\newcommand{\bea}{\begin{eqnarray}}
\newcommand{\ee}{\end{equation}}
\newcommand{\eea}{\end{eqnarray}}
\newcommand{\ds}{\displaystyle}
\newcommand{\high}[1]{\raisebox{10mm}{$#1$}}
\newcommand{\low}[1]{\raisebox{-1mm}{$#1$}}
\newcommand{\loww}[1]{\raisebox{-1.5mm}{$#1$}}
\newcommand{\lmn}{\mathop{\sim}\limits_{n\gg 1}}
\newcommand{\vpint}{\int\makebox[0mm][r]{\bf
--\hspace*{0.13cm}}}
\newcommand{\too}{\mathop{\to}\limits_{\beta\to\infty}}
\newcommand{\vp}{\varphi}
\newcommand{\vx}{{\vec x}}
\newcommand{\vy}{{\vec y}}
\newcommand{\vz}{{\vec z}}
\newcommand{\vk}{{\vec k}}
\newcommand{\vq}{{\vec q}}
\newcommand{\vpp}{{\vec p}}
\newcommand{\vn}{{\vec n}}
\newcommand{\vg}{{\vec \gamma}}
\newcommand{\kk}{K\bar{K}}

\begin{abstract}
In this manuscript we reply to the criticisms of our paper
[Eur. Phys. J. A 24, 437 (2005)] raised in a recent preprint by
Achasov and Kiselev [hep-ph/0606268] and demonstrate that all
their criticisms are completely irrelevant and misleading.
\end{abstract}
\maketitle

\begin{figure}[t]
\begin{center}
\begin{tabular}{cccc}

\epsfig{file=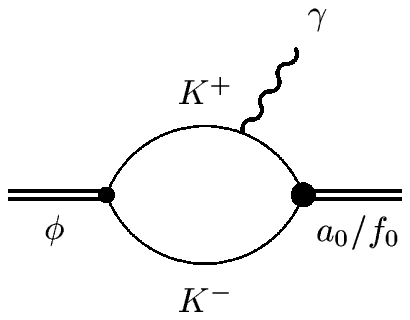,width=3.5cm}&
\raisebox{-6mm}{$\epsfig{file=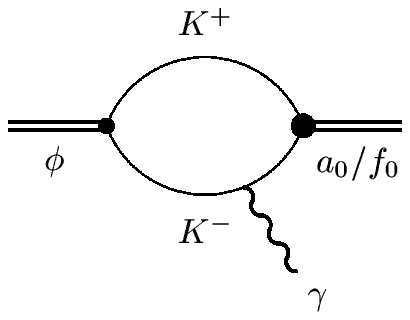,width=3.5cm}$}&
\epsfig{file=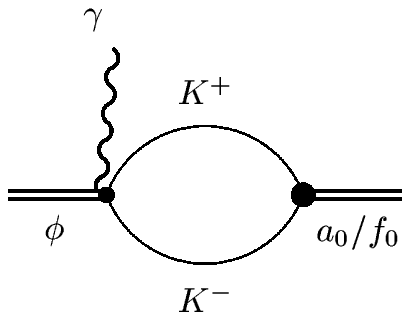,width=3.5cm}&
\epsfig{file=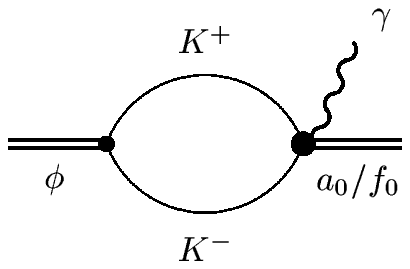,width=3.5cm}\\
(a)&(b)&(c)&(d)
\end{tabular}
\end{center}
\caption{Diagrams contributing to the amplitude of the radiative decay $\phi \to \gamma
a_0/f_0$.}
\end{figure}

In a recent paper \cite{mol} we considered the radiative decay
$\phi \to \gamma a_0/f_0$ in the molecular ($\kk$) model of the
scalar mesons ($a_0(980)$, $f_0(980)$). In particular, we showed that
there was no considerable suppression of the decay amplitude due to the
molecular nature of the scalar mesons. In addition, as a more general
result we demonstrated that, as soon as the vertex function of the
scalar meson is treated properly, the corresponding decay rates
become very similar to those for pointlike (quarkonia) scalar mesons,
provided reasonable values are chosen for the range of the interaction.
We also confirmed the range of order of $10^{-3}\div 10^{-4}$
for the branching ratio obtained in Refs.~\cite{Oset,Markushin,Oller}
within the molecular model.

Our results imply that differences in the concrete calculations of the
rate for $\phi \to \gamma a_0/f_0$ for a molecular and a quark ($q\bar q$ 
or $qq\bar q\bar q$) structure of the light scalar mesons can be only 
hidden in the effective coupling constants. In
either case --- regardless whether the scalars are molecules or not
--- the algebra how to calculate the rates is identical. Furthermore,
the form factors representing the involved ($\kk$) vertices should be
similar. In fact, we have argued that
the effective couplings extracted from the $\phi$ radiative
decays seem to be more consistent with those expected
for a molecular state or, at least, for a state containing 
a large admixture of the $K \bar K$ molecule. This is not
surprising since for the kaon loop mechanism of radiative decay 
to be operative, one needs the scalar effective coupling to 
be sizeable, which, in turn, requires a sizeable admixture of 
the $K \bar K$ in the scalar wavefunction, as demonstrated
in \cite{evi}. In other words, it means that the scalar 
radiative decay in the kaon loop mechanism proceeds via 
the molecular component of the wavefunction.  

As a reaction to our work a critical comment appeared \cite{AK}, where the
authors dispute our results and, specifically, where they claim
that our paper \cite{mol} is ``misleading''. We shall refute
the arguments of the authors of Ref. \cite{AK} below in detail one
by one.

But first, let us briefly recall the essentials of the formalism.
If the scalar meson is
considered to be point-like, the current describing the
transition $\phi\to\gamma S$ ($S$ = $a_0$ or $f_0$) is given by the
diagrams (a)-(c) of Fig.~1, and can be written as
\be
M_{\nu}=\frac{eg_{\phi}g_S}{2\pi^2im^2_K}I(a,b)
[\varepsilon_{\nu}(p\cdot q)-p_{\nu} (q\cdot \varepsilon)],
\label{loop}
\ee
where $p$ and $q$ are the momenta of the $\phi$ meson and
the
photon, respectively, $m_K$ is the kaon mass,
$g_\phi$ and $g_S$ are the $\phi K^+K^-$ and $SK^+K^-$
coupling constants,
$\varepsilon_\nu$ is the polarisation four--vector of the
$\phi$ meson,
$a=\frac{m^2_\phi}{m^2_K}$, and $b=\frac{m^2_S}{m^2_K}$ (we
deal here with stable scalars --- see the comment on nonstable scalars below).
The amplitude (\ref{loop}) is transverse, $M_{\nu}q_{\nu}=0$,
and it is proportional to the photon momentum. The form
(\ref{loop}) is well known, and the details can be found, for
example, in Refs. \cite{AI,Nussinov,Lucio,Bramon,CIK}. Specifically,
the total matrix element is finite if evaluated consistently with
gauge invariance and the integral
$I(a,b)$ entering Eq.~(\ref{loop}) remains finite in the
limit $a \to b$.

Now, if the scalar $S$ is treated as a non-point-like object, a
corresponding form factor is to be attached to the
$K^+K^-S$ vertex.
Note that the appearance of non-point-like vertices is in full
agreement with
the requirements of Quantum Field Theory (QFT). It can be derived
naturally as a solution to the Bethe--Salpeter equation in any
suitable QFT-inspired model for the interaction of kaons which is
able to bind a $\kk$ molecule. Thus it does not, by any means,
stem from ``$\ldots$ wrong interpretation of the result of calculation
in the point like case$\ldots$", as was claimed in Ref.~\cite{AK}.
Also an extra graph (graph (d) of
Fig.~1) should be added, which accounts for the additional flow
of the charge. The form for this extra contact vertex was
suggested in Ref.~\cite{CIK}, and is based on the minimal substitution
considerations. Indeed, let the scalar vertex $\varGamma$
depend on the total momentum of the scalar $P$ and on the
momentum $k$ of the $K^+$. Then, making the replacement $k_{\mu} \to
k_{\mu}-eA_{\mu}$, one obtains the ``interaction current"
\be
-e\frac{\partial \varGamma(k,P)}{\partial k_{\nu}} \ .
\label{contact}
\ee

On the other hand,
the authors of the preprint \cite{AK} repeatedly claim that
gauge invariance of the matrix element should be imposed
by hand, namely by a subtraction at the point $q=0$.
For example, in the paper \cite{AGS}, devoted to the
calculation of the $\phi \to \gamma S$ width in the
molecular model, it is written explicitly: ``$\ldots$the decay amplitude
$$
T(p,q)=M(p,q)-M(p,0),
$$
\be
M(p,q)=M_1(p,q)+M_2(p,q)+M_3(p,q),
\label{subtraction}
\ee
where $p$ and $q$ are the four-momenta of the $\phi$ meson
and photon respectively. ... The amplitudes $M_1$, $M_2$ and
$M_3$ correspond to the diagrams Fig.~1(a)-(c), respectively$\ldots$."

Let us start by stressing that,
generally speaking, both prescriptions (\ref{contact}) and
(\ref{subtraction}) fail to restore gauge invariance of
the amplitude. Indeed, the sum of the graphs depicted at Fig.~1(a)-(c) is given by
$M_{\nu}=eg_{\phi}g_S\varepsilon_{\mu}J_{\mu\nu}$
with
$$
J_{\mu\nu}=\int\frac{d^4k}{(2\pi)^4}(2k+2q-p)_{\mu}(2k+q)_{\nu}\varGamma(k,p-q)S(k)S(k-p+q)S(k+q)
$$
\be
+\int \frac{d^4k}{(2\pi)^4} (2k-p)_{\mu}(2k-2p+q)_{\nu}\varGamma(k,p-q)S(k)S(k-p+q)S(k-p)
\label{J}
\ee
$$
-2g_{\mu\nu}\int\frac{d^4k}{(2\pi)^4}\varGamma(k,p-q)S(k)S(k-p+q),
$$
where
\be
S(k)=\frac{1}{k^2-m^2+i0}
\ee
is the propagator of the $K$ meson, where here and in the following we
use the notation $m=m_K$ for simplicity.
Gauge invariance requires that the amplitude is transverse, while
\be
q_{\nu}J_{\mu\nu}=\int\frac{d^4k}{(2\pi)^4}
S(k)S(k-p)(2k-p)_{\mu}[\varGamma(k+q,p-q)-\varGamma(k,p-q)] \neq 0,
\label{qJ}
\ee
and both prescriptions (\ref{contact}) and (\ref{subtraction})
are obviously unable to repair this.

But also here the solution to this problem is well-known in the literature
(see, Ref.~\cite{BS}, where we list a few of 
the papers devoted
to this subject). In principle, one is to take into account the strong
interaction of the $\kk$ pair in the intermediate state, i.e. with the
interaction kernel responsible for the bound scalar state formation.
Inserting the photon line inside these graphs (the so-called
structure emission) restores gauge invariance.
However, in the
soft--photon limit, one can safely do without such complications.
Indeed, note that for soft photons (which is obviously the case for
the radiative decay under consideration), one may safely use the
current (\ref{contact}), so that, {\em in the prescription}
(\ref{contact}), the piece that violates gauge invariance
in Eq.~(\ref{qJ}) is cancelled by an extra
contribution to $J_{\mu\nu}$,
\be
-\int\frac{d^4k}{(2\pi)^4}S(k)S(k-p)(2k-p)_{\mu}\frac{\partial
\varGamma(k,p)}{\partial k_{\nu}},
\label{d}
\ee
coming from the
contact vertex (\ref{contact}) (through the contact graph depicted
in Fig.~1(d)). Therefore, {\em in the prescription (\ref{contact})},
the total current {\em is} transverse \cite{se} in the soft--photon
limit, i.e. up to corrections of the order $\omega/m$ with
$\omega$ being the photon energy.

Consider now the prescription (\ref{subtraction}). Gauge
invariance indeed requires the amplitude with neutral
particles in the initial and final state to vanish at $q=0$.
However, this alone {\em does not at all suffice to ensure gauge invariance},
as seen from (\ref{qJ}). In addition such a subtraction procedure {\em is not unique}.
The subtracted piece is only defined up to $q$--dependent contributions that vanish
in the limit $q=0$, but which might spoil the transversality of the transition current.
Thus, there are corrections in the order of $\omega/m$ too.
And, finally, the properly
calculated amplitude {\em must vanish at $q=0$ by itself, without extra subtraction,
finite or otherwise} \ftnote{1}{The case of light--by--light
scattering mentioned in \cite{AK} cannot be considered as a
counter-example: the amplitude calculated in the dimensional
regularisation scheme or in the background field method nicely vanishes automatically in the
limit $\omega \to 0$.}.
Thus, based on those considerations one can also put into question the entire
subtraction procedure suggested and advocated by one of the authors of the
preprint \cite{AK}. However, for the present discussion it is important to
note that, in the soft photon limit, the subtraction
prescription works and, as a matter of facts, {\em it provides results
identical to those given by the prescription (\ref{contact})}.
To show this, let us blindly use the prescription (\ref{subtraction}).
The expression (\ref{J}) for the transition current reads at $q=0$:
$$
J_{\mu\nu}(q=0)=\int \frac{d^4k}{(2\pi)^4}
(2k-p)_{\mu}(2k)_{\nu}\varGamma(k,p)S(k)S(k-p)S(k)
$$
\be
+\int \frac{d^4k}{(2\pi)^4} (2k-p)_{\mu}(2k-2p)_{\nu}
\varGamma(k,p)S(k)S(k-p)S(k-p)
\label{J0}
\ee
$$
-2g_{\mu\nu}\int \frac{d^4k}{(2\pi)^4}
\varGamma(k,p)S(k)S(k-p).
$$
If the Ward identity,
\be
2k_{\nu}S(k)S(k)=-\frac{\partial S(k)}{\partial k_{\nu}},
\label{byparts}
\ee
is used in the first term above and an integration by parts is performed
then the result reads
\be
J_{\mu\nu}(q=0)=-\int \frac{d^4k}{(2\pi)^4}
S(k)S(k-p)(2k-p)_{\mu}\frac{\partial
\varGamma(k,p)}{\partial k_{\nu}},
\label{equiv}
\ee
which is obviously identical to the expression (\ref{d}) found for
the other prescription. Therefore, we conclude that the criticism made in
Ref.~\cite{AK} regarding the alleged violation of gauge invariance
by the amplitude derived with the help of the prescription (\ref{contact})
is rather questionable.

After this lengthy introduction let us discuss the main
formula of our paper \cite{mol}. It is argued there (and
confirmed by actual calculations) that, since the
amplitude is finite even for the point-like limit,
the range of convergence of the integrals involved
{\em is defined only by the kinematics of the problem}. In
particular, if both masses, i.e. that of
the vector and of the scalar mesons, are close to the
$K \bar K$ threshold, the integrals
converge at $k_0\sim m$ and for nonrelativistic values of
the three--dimensional loop momentum $\vk$, $|\vk| \ll m$.
This allows us to perform a nonrelativistic reduction of the
amplitude in the rest-frame of the $\phi$ meson. The
contact scalar vertex is taken in the form (\ref{d}),
yielding for the individual graphs of Fig.~1:
\begin{eqnarray}
\nonumber J^{(a)}_{ik}=J^{(b)}_{ik}&=&-\frac{i}{m^3}\int\frac{d^3
k}{(2\pi)^3}
\frac{k_i(\vk-\frac{1}{2}\vq)_k\varGamma(|\vk-\vq/2|)}{[E_V-\frac{k^2}{m}+i0]
[E_S-\frac{(\vk-\vq/2)^2}{m}+i0]}, \\
J^{(c)}_{ik}&=&-\frac{i}{2m^2}\delta_{ik}\int\frac{d^3 k}
{(2\pi)^3}\frac{\varGamma(k)}{E_S-\frac{k^2}{m}+i0},\label{explint}
\\
J^{(d)}_{ik}&=&-\frac{i}{2m^2}\int
\frac{d^3k}{(2\pi)^3}\frac{k_ik_k}{E_V-\frac{k^2}{m}+i0}
\frac{1}{k}
\frac{\partial \varGamma(k)}{\partial k},\nonumber
\end{eqnarray}
where $E_V=m_V-2m$, $E_S=m_S-2m$. Integration by parts was
performed for the last integral, yielding
\be J^{(d)}_{ik}= \frac{i}{2m^2}\delta_{ik} \int
\frac{d^3 k}{(2\pi)^3}
\frac{\varGamma(k)}{E_V-\frac{k^2}{m}+i0}+
\frac{i}{3m^3}\delta_{ik} \int \frac{d^3 k}{(2\pi)^3}
\frac{k^2\varGamma(k)}{(E_V-\frac{k^2}{m}+i0)^2} \ .
\label{nbyparts}
\ee

Surprisingly, it is this procedure of integration by parts which
caused the main criticism in Ref.~\cite{AK}. Indeed, it is difficult
to understand why the authors of Ref.~\cite{AK} overlooked the simple
fact that, by virtue of Eq.~(\ref{equiv}), even with the subtraction
procedure (\ref{subtraction}) one would arrive {\em at
the very same result} (\ref{nbyparts}). Schematically, this looks as
follows (for the sake of brevity we omit indices). Let us start with
the form (\ref{explint}) and denote the sum of the real parts of the
integrals $J^{(a)}+J^{(b)}+J^{(c)}$ in Eq.~(\ref{explint}) as $J(q)$.
Let us follow for a moment the authors of Ref.~\cite{AK} and introduce
the cut-off $k_0$ to compute the integrals $J(q,k_0)$ and $J^{(d)}(k_0)$.
But it is important to realize that such a sharp cut-off obviously breaks
gauge invariance -- something which the authors of the preprint \cite{AK}
have either overlooked or tacitly ignored.
Then, as a consequence, the sum $J(q,k_0)+J^{(d)}(k_0)$ clearly exhibits a
drastic dependence on the cut-off: the limiting value
$J(q,\infty)+J^{(d)}(\infty)$ is not reproduced with the values of $k_0$
up to several GeV (see Figs.~2, 3 of Ref.~\cite{AK}).
This behaviour is demonstrated in Ref.~\cite{AK} and it is advertised
as being the collapse of the approach suggested in Ref.~\cite{mol}.
However, one can easily see that
the sum $J(q,\infty)+J^{(d)}(\infty)$ vanishes at $q=0$, while
the sum $J(q,k_0)+J^{(d)}(k_0)$ does not if $J^{(d)}$ is taken in the
form (\ref{explint}). Thus, as already mentioned before, such a
cut-off imposed by brute force {\em does not respect gauge
invariance} and consequently, and in line with the arguments of one of
the authors of the preprint \cite{AK}, calls for the subtraction:
\be
J^R(q,k_0)=J(q,k_0)+J^{(d)}(k_0)-[J(q,k_0)+J_{(d)}(k_0)]_{|q=0}=
J(q,k_0)-J(0,k_0).
\ee
If then $J^{(d)}$ is chosen as defined in Eq.~(\ref{nbyparts})
one obtains $J(0,k_0)=J^{(d)}(k_0)$ and,
thus, the procedure of integration by parts after performing
nonrelativistic reduction is {\em perfectly justified} even in the
subtraction scheme (\ref{subtraction}) advocated by one of the
authors of the preprint \cite{AK}.
The regularised integral $J^R(q,k_0)$ {\em converges rapidly and
for low virtual momenta of the kaons} -- which follows both from our
findings and from the one of the preprint \cite{AK} (see Fig.~4 of
Ref.~\cite{AK}).
In this context, let us note that the authors of Ref.~\cite{AK}
have deliberately forgotten to mention that, for realistic values
for the range of the interaction, $\beta$, i.e. $\beta \sim 0.6$ GeV or
larger, the result for the matrix element is actually very close
to the one of the point-like theory.

Therefore, we conclude that there are no flaws in the
derivation presented in Ref.~\cite{mol}. This is in a drastic contrast
to the statement of Ref.~\cite{AK}, where it is claimed that high kaon
virtualities are dominating the radiative transition
amplitude in the point-like limit. Note that, if this would be indeed
the case, then the kaon loop mechanism for the $\phi$ radiative decay
must be {\em completely irrelevant}: there are no infinitely
compact states in nature. Besides, the convergence of an integral for
values of the loop momentum much larger than all the natural scales
present in the problem looks rather suspicious from a physical point
of view (and is also aesthetically unacceptable).

A further claim by the authors of Ref.~\cite{AK} is that
``$\ldots$ there is another sloppily build place in Ref.~\cite{mol}.
The authors of Ref.~\cite{mol} calculate $A(\phi(p \to \gamma  S(p')$ at $q=0$,
$p^2=m_{\phi}^2$ and $(p')^2=m_S^2~(m_S-2m<0$), that is, at $p \neq
p'+q.\ldots$" (see p.~7 (reference [6]) of Ref.~\cite{AK}). This is simply
{\em not true} and it remains completely unclear to us what made them to draw
that conclusion. The amplitude was calculated
retaining the full dependence on $\vq$ in Eq.~(\ref{explint}). It turned out,
however, that, due to the inequality
\be m(E_V-E_S) \gg {\vq}{\,^2}\approx (E_V-E_S)^2,
\ee
one may safely omit {\em numerically} the terms containing $\vq$ in the
expression (\ref{explint}). This does not mean that the
calculations were performed under the assumption $p \neq p'+q$.\ftnote{2}{Strictly
speaking, it is not legitimate to keep
subleading terms in $q$, as both prescriptions (\ref{contact}) and
(\ref{subtraction}) maintain gauge invariance only in the soft
photon limit, as was discussed above.}

Yet another wrong statement of Ref.~\cite{AK} concerns the problem
of unstable scalars. It remains unclear how it is related
to the ``question of principle" raised at p.~7 of Ref.~\cite{AK}, but
the actual situation is as follows. An attentive and unbiased reader
would have noticed that, while in the main body
of our paper \cite{mol} the case of the stable scalar with
$m_S-2m<0$ is considered, the Appendix is entirely devoted to
the case of the unstable scalar which can decay into a pair of
light pseudoscalars $P_1P_2$. Contrary to the claims of Ref.~\cite{AK}, the
$P_1P_2$ invariant mass distribution is calculated, both in the differential
form and also integrated over the near-threshold mass region,
$900$ MeV$<M<m_{\phi}$. The results presented in \cite{mol}
appear to be {\em robust against the inclusion of the finite width of the scalar}.

To summarise, {\em it is the statement of the preprint \cite{AK} to be completely
misleading}. There are no reasons at all to believe that the
radiative decays of the $\phi$ meson into scalars would
allow to discriminate between molecules and compact states.

Finally, let us reiterate the main results Ref. \cite{mol}.
There, we have shown that there is no
appreciable suppression of the $\phi\to\gamma S$ width in the
molecular model for the scalar mesons, provided reasonable values
are chosen for the range of the interaction, $\beta \sim 0.6-0.8$
GeV. In the molecular model this range is defined by the mass of the
lightest meson participating in the exchange force, that is by the
$\rho$ meson. In recent kaon-loop calculations \cite{Oset,Markushin,Oller}
where the scalars are
considered as dynamically generated states, that is, as molecules,
a range of the interaction of similar order was used, and a good
description of the data on the $\phi$ radiative transitions was
achieved. If the scalars are treated as $q \bar q$ states, the range
of interaction can be estimated from the range of the vertex
responsible for the strong decays of the mesons. This can be deduced,
for example, within the $^3P_0$ model for pair creation \cite{Barnes}
to be $\beta \sim 0.8$ GeV. Therefore, it is not possible to
distinguish between a molecular or a $q \bar q$ nature of the
$a_0(980)$ and $f_0(980)$ from data on $\phi$ radiative decays
based on the value of the relevant interaction range because they
are essentially the same.
The only real difference in the concrete calculations of the rate for
$\phi \to \gamma a_0/f_0$ for a molecular and a quark ($q\bar q$ or
$qq\bar q\bar q$) structure of the light scalar mesons is 
hidden in the effective coupling constants. 
The ``compactness'' advocated in Ref.~\cite{AK} 
as a criterion for the discrimination is completely irrelevant
for the $\phi$ radiative decay. 

\begin{acknowledgments}
The authors are grateful to N. Achasov and A. Kiselev for finally giving credit and bringing 
attention to their manuscript \cite{mol}.
This research was supported by the grants RFFI-05-02-04012-NNIOa, DFG-436 RUS 113/820/0-1(R), NSh-843.2006.2, and
NSh-5603.2006.2, by the Federal Programme of the Russian Ministry of
Industry, Science, and Technology No. 40.052.1.1.1112, and by the Russian
Governmental Agreement N 02.434.11.7091. A.E.K. acknowledges also partial support
by the grant DFG-436 RUS 113/733.
\end{acknowledgments}


\begin{thebibliography}{99}
\bibitem{se} One might think that the
current (\ref{d}) given by minimal subtraction is not
unique, and some transverse contribution can be added. It is
not the case in the soft photon limit. As soon as the strong
interaction problem is defined, the structure emission term
should reduce to (\ref{contact}) for small $q$, though the
actual pattern of this reduction is model-dependent. For
example, expression for structure emission term in the
Bethe-Salpeter equation framework explicitly exhibits
this feature. The details can be
obtained from the authors by request.
\end{thebibliography}
\end{document}